# Lifetimes of rogue wave events in direct numerical simulations of deep-water irregular sea waves

**Anna Kokorina** [1], **Alexey Slunyaev** [1,2*]

[1] Institute of Applied Physics, 46 Ulyanovs, Box-120, Nizhny Novgorod 603950, Russia
[2] National Research University-Higher School of Economics, 25 B. Pechorskaya Street, Nizhny Novgorod 603950, Russia
* Correspondence: a.sergeeva@appl.sci-nnov.ru

**Abstract:** The issue of rogue wave lifetimes is addressed in this study, which helps to detail the general picture of this dangerous oceanic phenomenon. The direct numerical simulations of irregular wave ensembles are performed to obtain the complete accurate data on the rogue wave occurrence and evolution. The simulations are conducted by means of the HOS scheme for the potential Euler equations; purely collinear wave systems, moderately crested and short-crested sea states have been simulated. We join instant abnormally high waves in close locations and close time moments to new objects, rogue events, what helps to retrieve the abnormal occurrences more stably and more consistently from the physical point of view. The rogue wave event probability distributions are built based on the simulated wave data. They show the distinctive difference between rough sea states with small directional bandwidth on the one part, and small-amplitude states and short-crested states on the other part. The former support long-living rogue wave patterns (the corresponding probability distributions have heavy tails), though the latter possess exponential probability distributions of rogue event lifetimes and produce much shorter rogue wave events.

---

## 1. Introduction

Rogue (or freak) waves are one of the most intriguing natural phenomena in the sea which has received much attention in the recent years. Today there is no doubt that waves exceeding the significant wave height in two-three times and even more do happen in the Ocean [Kharif et al, 2009]. However, do abnormally high waves belong to the class of waves driven by new, so far unaccounted, physical mechanisms, or they are rare events caused by the co-phased superposition of stochastic sea waves – is still an issue of scientific debates [Haver & Andersen, 2000; Zakharov et al, 2006; Onorato et al, 2013; Christou & Ewans, 2014; Fedele et al, 2016]. Though the amount of instrumental in-situ wave registrations seems to be huge, the problems of justification of reliability of these data and of performing accurate statistical analysis are challenging. The direct numerical simulation of primitive hydrodynamic equations is considered nowadays as an appropriate approach to avoid the drawbacks of in-situ experiments and to obtain precise data on realistic waves in fully controllable conditions, e.g. [Tanaka, 2001; Chalikov et al, 2005; Toffoli et al, 2008; Xiao et al, 2013; Bitner-Gregersen et al, 2014; Ducrozet et al, 2016; Brennan et al, 2018]. The availability of the full wave data in the numerical simulations gives clues about the questions which could hardly be answered in the near future based merely on the experimental observations.

The lifetime of extreme events is one of the wave characteristics which is difficult to measure in-situ, though may be straightforwardly estimated based on the direct numerical simulations. According to people's observations, the lifetimes of terrifying rogue waves can amount to a few minutes or less (see e.g. in [Kharif et al, 2009]). For example, walls of water were seen for "a couple of minutes" during the accidents with the liner Queen Elisabeth II (1995) and with a Statoil platform

Veslefrikk B (1995) [Haver & Andersen, 2000]. The notorious knowledge that only a single huge wave is most frequently reported by eyewitnesses of rogue waves may be related to the fact that the observers (or the measuring probes) are located in single points and cannot follow the travelling wave patterns. Hence the typical duration of extreme wave events remains under a veil of mystery so far.

The characteristic life times of deterministic wave groups evolving under the action of effects of the wave dispersion or the nonlinear self-modulation were concerned in [Pelinovsky et al, 2011; Slunyaev & Shrira, 2013]. Based on the laboratory experiments with focusing wave groups, the life times of 1-3 minutes were estimated in [Shemer et al, 2007]. The issue of life times of rogue waves which occur in numerical simulations of irregular unidirectional water waves was particularly addressed in the researches [Sergeeva & Slunyaev, 2013; Slunyaev et al, 2016].

In this work we will follow the most popular definition of a rogue wave as the wave which exceeds the significant wave height at least twice,

$$AI \equiv H/H_s \geq 2\,, \tag{1}$$

where $H$ is the wave height and $H_s$ is the significant wave height; $AI$ means the amplification index. As in deep water wave envelopes propagate with the velocity twice slower than individual waves, the maximum wave height within a group oscillates in time. Hence, depending on the wave phase the particular wave pattern may be referred to the rogue wave class or not (see the discussion in [Dysthe et al, 2008]). In [Sergeeva & Slunyaev, 2013; Slunyaev et al, 2016] *rogue waves* which occur in close locations in space and at near time instants were combined together and referred to *rogue events*. There the lifetimes of rogue events were found as long as tens of wave periods (up to 60), what results in maximum 10 minutes if the waves possess periods of 10 s. An extremely long-living event was presented in the paper [Slunyaev & Kokorina, 2017], where an intense wave group occasionally emerged in the field of random waves, and then was persisting for about 200 wave periods within the strongly nonlinear numerical simulation of collinear waves. The sea state was relatively moderate being a realization from the series $A_0$ in Table 1, characterized by the JONSWAP spectrum with the peakedness $\gamma$ = 3, the dominant wave period $T_p$ = 10 s and the significant wave height $H_s$ = 3.5 m (a realization from the series $A_0$ in Table 1 below). In a very recent numerical simulations of directional deep water waves [Fujimoto et al, 2019] the maximum registered lifetime of rogue events was limited by 30 wave periods.

In the present paper we consider lifetimes of rogue wave events in statistical sense. The wave data are obtained in direct numerical simulations of deep water waves governed by the potential hydrodynamic equations restricted to the accounting for the four-wave nonlinear interactions. The simulated sea states are described by the JONSWAP spectrum with three degrees of directional spreading. The results obtained in the numerical simulations of nonlinear waves are also compared with the linear framework. The description of the problem setting and of the methodology is given in Sec. 2. The case of unidirectional waves is discussed first in Sec. 3. Results of the numerical simulations of directional waves are collected in Sec. 4, while the conclusions are collected in the subsequent section.

## 2. Description of the approach

In this work the data on evolving water surfaces are obtained in numerical simulations of directional deep-water waves within the Euler equations for potential flows. The equations are solved with the help of the High Order Spectral Method [West et al, 1987] which uses the decomposition of the velocity potential in the vicinity of the water rest level truncated to the third order, $M$ = 3. This approximation corresponds to the consideration of up to the four-wave nonlinear interactions. Such simplification helps to reduce the simulation time and is commonly used as the four-wave interactions dominate in the deep water conditions. The integration in time is performed after splitting of the governing equations into the linear and nonlinear parts. The linear part is solved at each time step with the help of the analytical solution, though the nonlinear part is solved with the help of the 4-order Runge-Kutta method. A weak dissipation was introduced to the scheme by virtue of the low-pass spectral filter, similar to [Xiao et al 2013]. The filter helps to reduce the effect of occasional wave breaking and to stabilize the code. Besides smoothing of too steep waves, the spectral filter causes noisy small-scale perturbations if compared to the original solution of the non-dissipating equations, see a thorough consideration in [Slunyaev & Kokorina, 2019].

The initial conditions for the numerical simulations were specified according to the JONSWAP spectrum with the peak wave period $T_p \approx 10$ s, peakedness $\gamma \approx 3$ and different sets of the significant wave heights, $H_s$, and of the directional wave spreading, $\theta$, specified by the function $D(\chi)$,

$$D(\chi) = \begin{cases} \frac{2}{\theta}\cos^2\left(\frac{\pi\chi}{\theta}\right), & |\chi| \le \frac{\theta}{2} \\ 0, & |\chi| > \frac{\theta}{2} \end{cases}. \quad (2)$$

Parameters of the simulations are given in Table 1. The low sea states are denoted with the letter A, while the rough sea states – with E. The subscripts in the numerical simulation coding give the value of the directional spreading, $\theta = 0$, $\theta = 12°$ or $\theta = 62°$.

**Table 1.** Parameters of the numerical experiments

| Exp. code | $\gamma$ | $T_p$, s | $\theta$, grad | $H_s$, m |
|---|---|---|---|---|
| $A_0$ | 3 | 10 | 0 | 3.5 |
| $E_0$ | 3.3 | 10.5 | 0 | 7 |
| $L_{12}$ | 3 | 10 | 12 | linear |
| $A_{12}$ | 3 | 10 | 12 | 3.5 |
| $E_{12}$ | 3 | 10 | 12 | 6 |
| $L_{62}$ | 3 | 10 | 62 | linear |
| $E_{62}$ | 3 | 10 | 62 | 7 |

Each realization of directional waves was generated at the moment $t = 0$ in the area of 50 by 50 dominant wave lengths $\lambda_p$ (what is about 8 km by 8 km) according to the linear wave theory. Here the wave period and its length are assumed related by the linear dispersion relation, $2\pi/T_p = (2\pi g/\lambda_p)^{1/2}$, $g$ is the gravity acceleration. The double-periodic boundary conditions are used.

The nonlinear adjustment method suggested in [Dommermuth, 2000] was applied for the first 20 wave periods, $0 < t < 200$ s. During this preliminary stage the nonlinear terms of the simulated equations were being slowly turned on. It allows the initially linear waves to transform adiabatically into nonlinear wave solution and to avoid the generation of spurious high-frequency components. The experiments marked in the column $H_s$ of Table 1 as 'linear' were simulated according to the linear equations of hydrodynamics with the full linear dispersion; the corresponding simulations are coded with the letter L.

After the preliminary stage, the waves were simulated in the nonlinear regime for the next 120 dominant wave periods, $200\text{ s} < t < 1400$ s. The simulated wave fields were stored each 0.5 s (i.e., the database contains 20 snapshots of the surface per one wave period). The spatial resolution used in the present study is 1024 × 1024 mesh points, which results in about 20 points per one dominant wave length in the longitudinal and the lateral directions (see the discussion of sufficiency of these parameters in [Slunyaev & Kokorina, 2019]). For every directional wave condition 6-7 realizations were simulated, what yields datasets of about 400 km$^2$ simulated for 20 minutes (one realization gives 50 waves by 50 waves by 120 periods).

Characteristics of the simulated wave fields, such as the dominant wavenumber and the spectral width, do slowly evolve in the course of the simulation for $200\text{ s} < t < 1400$ s. In general, the mean wavenumber and the mean frequency slightly grow, though their peak values decrease; the wave spectra for wavenumbers, frequencies and angles become wider. The effect of increasing of the wave directionality may be easily seen by an eye in Figure 1a, where the snapshots of the surfaces at $t = 200$ s and $t = 1400$ s are shown for the rough sea case with initially narrow angle spectrum. The evolution of spectrum is much less obvious for the situation of shorter-crested waves shown in

Figure 1b. The surface displacement variance remains approximately constant during the simulations as the energy is conserved with very good accuracy. The relative error of conservation of the total energy in the period 200 s < $t$ < 1400 s is less than $1{\cdot}10^{-3}$.

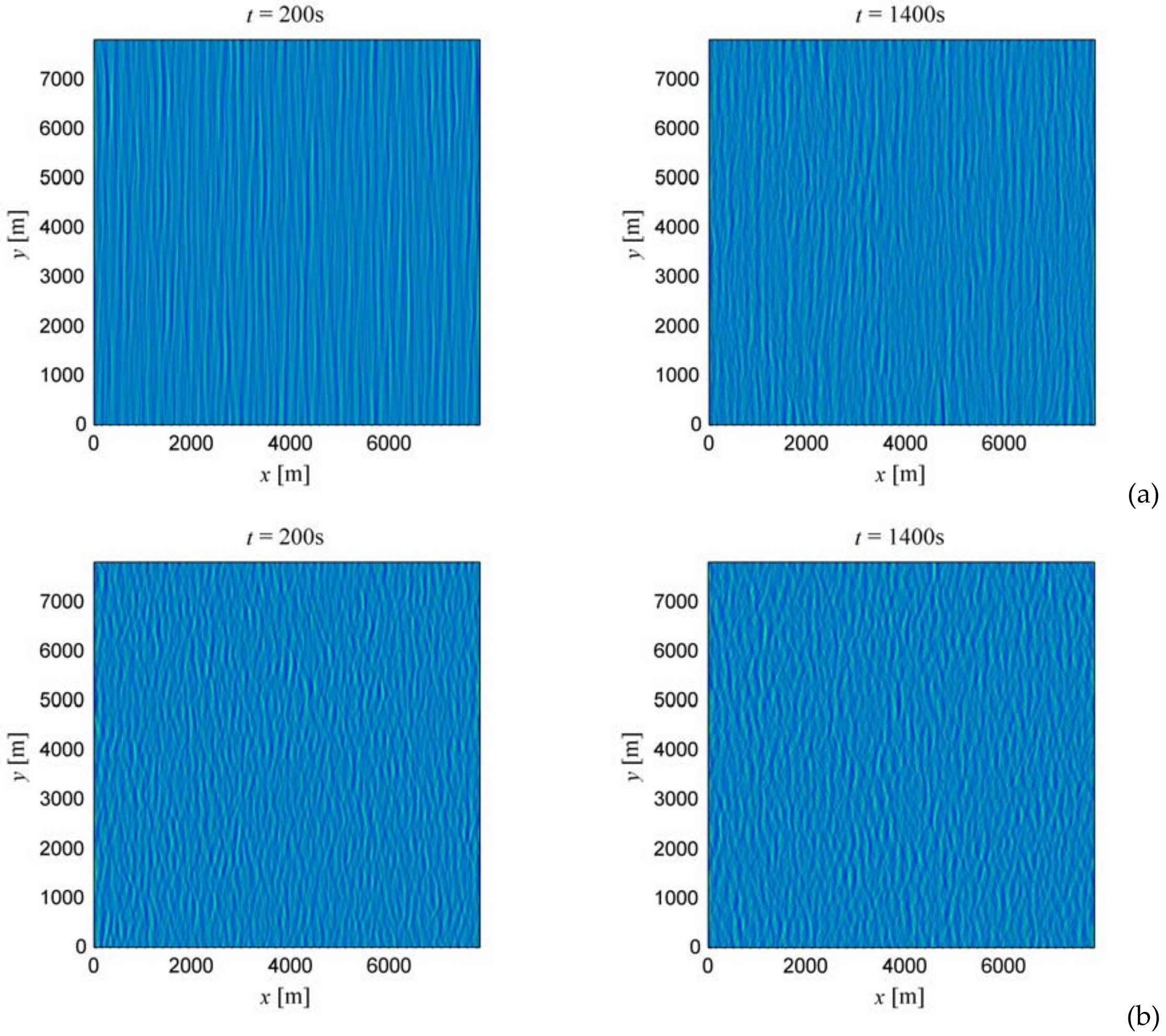

**Figure 1.** Examples of the simulated sea surfaces: (**a**) series $E_{12}$ and (**b**) series $E_{62}$.

The period of 20 min (120 wave periods) corresponds to the conventional time of the sea wave statistical quasi-stationarity and it is significantly longer than used in some numerical simulations (e.g., 50 periods in [Fujimoto et al, 2019]). This circumstance is twofold. On the one hand, the parameters of wave systems drift apart from the initial condition, and the averaging along this period is strictly speaking not accurate from the viewpoint of the statistical theory. On the other hand, the transition from the artificially prescribed initial condition is not limited by the 20-period nonlinear adjustment, which we perform following [Dommermuth, 2000]. A longer transition stage exists, $\sim(k_p\sigma)^{-2}T_p$, when wave groups of natural shapes appear [Slunyaev, 2010; Slunyaev & Sergeeva, 2011], hence the simulated sea states for first tens of wave periods may be still different from the natural condition. Therefore using a longer simulation can be advantageous.

The further processing is performed with the fields of the surface elevation, $\eta(x, y, t)$, where $x$ is the coordinate along the dominant wave propagation, and $y$ is the transverse coordinate. The surface at each instant $t$ is sliced into 128 cuts parallel to the $Ox$ axis (hence the cuts are spaced at the distance about $0.4\lambda_p$). The obtained in such way space series are about 8 km length each, and periodic. The space series are analyzed locating the wave crests with the use of the zero-crossing approach. The wave height associated with each of the crests is defined as the maximum vertical distance between the crest and preceding and following wave troughs. The significant wave height is evaluated for the space series through the variance, $H_s = 4\sigma$. This parameter (particular for every cut at each time instant) is used to select rogue waves according to the criterion (1). In this way rogue waves for the entire simulated area 8 km by 8 km are detected within the simulated time frame of 20 min.

The set of retrieved *rogue waves* is then reorganized into *2D rogue events*. To this end the fields $\eta(x, t)$ are analyzed for each longitudinal cut according to the same procedure as in [Sergeeva & Slunyaev, 2013; Slunyaev et al, 2016], i.e. two rogue waves belong to one 2D rogue event if the differences between their longitudinal coordinates $x - ct$ and times $t$ are less than $m\lambda_p$ and $mT_p$ correspondingly; their transverse coordinates $y$ coincide. Here the co-moving with the group wave velocity reference is used and the periodicity of the domain along $Ox$ is taken into account. In this work we consider three values of $m$, $m = 1$, $m = 2.5$ and $m = 4$.

At the next stage *3D rogue events* are picked out on the basis of the 2D rogue events revealed in the longitudinal cuts. Two *rogue waves* belong to one *3D rogue event* if the differences in their coordinates $x - ct$, $y$ and times $t$ are less than $m\lambda_p$, $m\lambda_p$ and $mT_p$ correspondingly. In the situations with narrow angle spectra ($L_{12}$, $A_{12}$, $E_{12}$) the rogue wave patterns have large lateral size and hence the 3D rogue events are much less in number than the corresponding 2D events.

In the experimental condition the set of 2D rogue events could be collected if one follows the evolution of sea waves along a straight line (for example, having an array of wave gauges along the direction of wave propagation). The 2D rogue events were previously analyzed in the strongly nonlinear simulations of strictly collinear waves reported in [Sergeeva & Slunyaev, 2013; Slunyaev et al, 2016; Slunyaev & Kokorina, 2017]. The data from those simulations are used in the present work (see series $A_0$ and $E_0$ in Table 1; they are studied in Sec. 3) to compare the dynamics of collinear and directional waves.

The set of 3D rogue events comprises the information on natural objects of the sea surface dynamics – the energetic patterns which manifest themselves through large individual waves. The extreme waves appear from time to time and disappear for a little due to the strongly irregular and transient character of sea waves. We assume that a rogue event lasts from the moment when the first rogue wave occurs till the instant of the last rogue wave, which belong to the given event.

Finally, we refer the very recent paper [Fujimoto et al, 2019] where rather similar analysis of rogue wave lifetimes was performed based on the direct numerical simulations by the HOSM with different orders of nonlinearity accounting for up to 4-wave interactions ($M$ = 1, 2, 3). They considered two sea states registered in the Pacific Ocean, with narrow and broad directional spectra. Compared to our approach, the main differences are the following. We perform zero-crossing analysis and seek for rogue waves in space series, not time series. We combine rogue waves into 3D events using a different method, from larger areas in time and space, and taking into account the wave drift due to the group velocity. The 2D rogue events were not considered in [Fujimoto et al, 2019].

## 3. Rogue wave lifetimes in the simulations of unidirectional waves

In this section we analyze the dataset of simulations of irregular strictly planar waves accumulated in our previous works [Sergeeva & Slunyaev, 2013; Slunyaev et al, 2016; Slunyaev & Kokorina, 2017], see series $A_0$ and $E_0$ in Table 1. In this case the procedure of selection of the rogue events is equivalent to collecting 2D rogue wave events as described above. The exceedance probability distributions of rogue event lifetimes are plotted in Fig. 2, where the probability is calculated according to the formula

$$P(T_n) = \frac{n}{N_{ev} + 1}, \qquad n = 1, 2, \ldots, N_{ev}. \tag{3}$$

In (3) $T_n$ are the lifetimes of the total number of $N_{ev}$ rogue wave events. The stability of estimation of the probability we characterize with the help of the standard deviation $\Delta P$ (similar to [Tayfun & Fedele, 2007]),

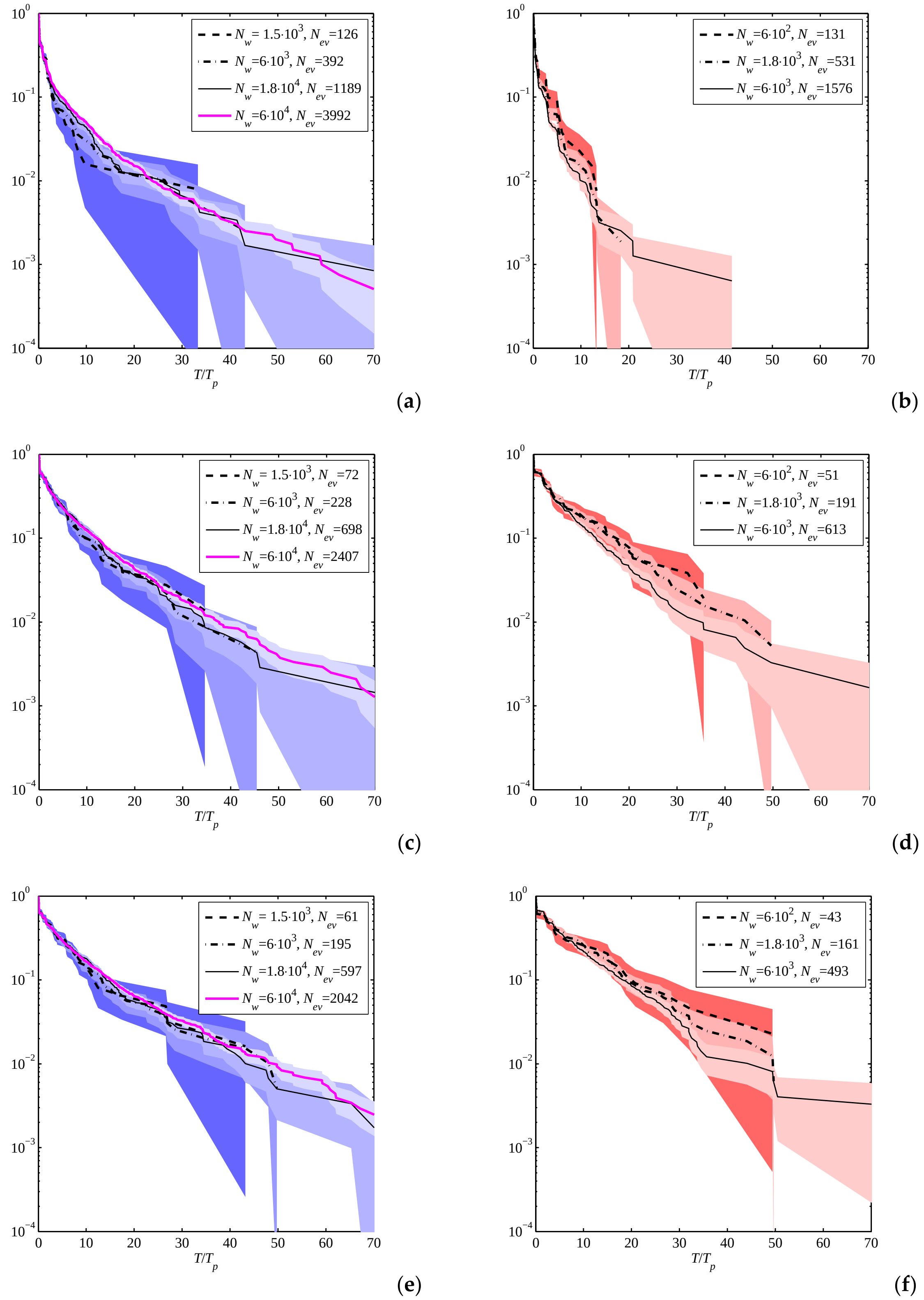

**Figure 2.** Exceedance probability distributions for rogue event lifetimes $P$ (lines) and the corresponding confidence intervals $P \pm \Delta P$ (filled areas, from deep to light colors correspond to from small to large statistical ensembles) for different databases. The legends give the number of the simulated for 1200 s waves, $N_w$, and the total number of found rogue wave events, $N_{ev}$. (**a**) series $A_0$, $m$ = 1; (**b**) series $E_0$, $m$ = 1; (**c**) series $A_0$, $m$ = 2.5; (**d**) series $E_0$, $m$ = 2.5; (**e**) series $A_0$, $m$ = 4; (**f**) series $E_0$, $m$ = 4.

$$\Delta P(T_n) = \frac{1}{N_{ev}+1}\sqrt{n\frac{N_{ev}-n+1}{N_{ev}+2}}, \qquad n = 1,2,\ldots,N_{ev}. \tag{3}$$

While the lines in Fig. 2 plot the values of $P$, the colors show the ranges $P \pm \Delta P$ for the given number of events $N_{ev}$.

The issue of necessary amount of data which is sufficient to distinguish the difference between extreme (and rare) event probabilities at different conditions is not obvious. In each of the Fig. 2a-f we plot the distributions for different volumes of statistical ensembles to check the convergence of the probabilistic description. The numbers $N_w$ in the legend denote the approximate numbers of waves in the sequences which were simulated for the period of about $120T_p$ and which compose the statistical ensembles. This approach of indirect verification of the probabilistic description does not always serve well as we emphasized in [Slunyaev & Kokorina, 2017], where very rare events were shown to be able to influence the probability distribution for relatively frequent events.

When the volume of the statistical ensemble grows with $N_w$, rarer and more extreme events occur, and correspondingly the curves of $P(T)$ generally continue to larger values of lifetimes $T/T_p$ and to smaller probabilities (down to $1/N_{ev}$). Though the curves $P(T)$ for different ensembles generally agree, noticeable discrepancies may be found not only in the tails of the distributions, but also at the levels of relatively frequent events (e.g., the solid curves in Fig. 2d,f). The confidence intervals shrink with the growth of $N_w$, they are shown with colors in Fig. 2, where lighter colors correspond to larger wave ensembles. The dependences plotted with the same line styles in the left and right sides of each row in Fig. 2 correspond to approximately similar numbers of $N_{ev}$. The overall ensemble in series $A_0$ consists of about $6\cdot10^4$ wave lengths simulated for 120 wave periods, i.e., $7\cdot10^6$ wave-periods; the corresponding curves are plotted in Fig. 2a,c,e with the magenta lines. The number of simulated realizations in the series $E_0$ is about 10 times smaller than in the series $A_0$, thus, the probability distribution for the series $A_0$ is expected to be more reliable.

It is not obvious a priori, which value of the parameter $m$, used when combining instant rogue waves into rogue events, is optimal. A too small value of $m$ would lead to consideration of one focused wave pattern as many independent extreme events due to the wave irregularity; a too large value may cause joining of physically independent events into one. In three rows of Fig. 2 we show the results for three values of $m$, from 1 to 4. In Fig. 2a,b rogue wave events are collected from the set of instant rogue waves with the allowed differences in their coordinates, $x - ct$, and times, $t$, no more than $\lambda_p$ and $T_p$ correspondingly ($m = 1$). This choice of the arrangement parameter $m$ was used in [Fujimoto et al, 2019]. Larger differences, $2.5\lambda_p$ and $2.5T_p$, were admitted in our works [Sergeeva & Slunyaev, 2013; Slunyaev et al, 2016], Fig. 2c,d correspond to this choice.

Note the drastic difference when comparing Fig. 2a versus Fig. 2c and Fig. 2b versus Fig. 2d. The difference between the distributions for the series $E_0$ for $m = 1$ and $m = 2.5$ is much more pronounced than that for the series $A_0$. The number of rogue events when $m = 1$ is larger, though their lifetimes are shorter; this effect is stronger for the steeper series $E_0$. When $m = 1$, the ratios $N_w/N_{ev}$ are about 1 800 wave-periods per rogue event in the series $A_0$ and about 450 wave-periods per rogue event in the series $E_0$. The distribution for the steeper sea state (Fig. 2b) is obviously below the one for the moderate steepness (Fig. 2a).

When $m = 2.5$ (Fig. 2c,d), the distributions look rather similar for the conditions of moderate (series $A_0$, Fig. 2c) and strong (series $E_0$, Fig. 2d) nonlinearity; the maximum lifetimes are about $70T_p$ in both the series. At the same time, the fraction of the number of simulated waves to the number of rogue events, $N_w/N_{ev}$, is remarkably different from the case $m = 1$: about 3 000 wave-periods per rogue event in the series $A_0$ and about 1 200 wave-periods per rogue event in the series $E_0$.

Even larger value of $m = 4$ was used to see the effect of this parameter on the distribution of rogue wave lifetimes, see Fig. 2e,f. Though the distributions alter when compared to Fig. c,d this difference is moderate, the ratios $N_w/N_{ev}$ are about 3 500 and 1 500 wave-periods per rogue event respectively for the series $A_0$ and $E_0$.

Bearing in mind the strong modification of the probability distribution for the simulations $E_0$ when $m$ changes from 1 to 2.5 and its smooth correction when $m$ grows further to the value of four, we assume that the value of $m = 1$ is too small for the simulations $E_0$. In general, small values of $m$ are beneficial for the statistical study as the number of events decreases when $m$ grows and hence the volume of the statistical ensemble diminishes. Thus, we consider the value of $m = 2.5$ optimal for the conditions of the simulations $A_0$ and $E_0$.

The dependencies $P(T)$ in Fig. 2c,d approximately follow the exponential laws with somewhat heavier tails. We note the stepped dependencies of $P(T)$ in the interval of short life-times for all the ensembles of the series $E_0$ (better seen in Fig. 2b); they represent local plateaux most visible at the lifetimes $T \approx 2T_p$ and $T \approx 4T_p$. Most likely these peculiarities are related to the group structure of waves and the difference between the phase and group velocities of the waves. The observed peculiarity corresponds to a low probability of wave groups consisting of about one and two individual waves correspondingly (in the spatial domain).

For the series $A_0$ with the largest ensemble, based on Fig. 2c one may say that the probability distributions seem to converge for the ensembles of a thousand of rogue events and more, $N_{ev} \geq O(10^3)$. Hence the statistical ensembles used for the study in this and the next sections are marginally sufficient or may be even too scarce to study the trends of the probability distributions with the minimum probability of the order of $10^2$.

## 4. Rogue wave lifetimes in directional fields

In this section the results of 3D simulations of waves with two directional spreads (see the description in Sec. 2) are discussed and compared with the results of simulations of planar waves discussed in the previous section. In Fig. 3 we plot the exceedance probability distributions of the 2D rogue wave event lifetimes (Fig. 3a) and of 3D events (Fig. 3b) for the conditions listed in Table 1 and the arrangement parameter $m = 1$. The same dependencies for $m = 2.5$ and $m = 4$ are plotted in Fig. 3c,d and Fig. 3e,f respectively. The difference between the plots for different $m$ is obvious. According to the analysis of the lifetime dependencies performed in Sec. 3 for the unidirectional waves, the choice $m = 1$ is not appropriate, while $m = 2.5$ is probably optimal. The lifetimes in the linear simulations of directional waves ($L_{12}$, $L_{62}$) posses similar distributions which slightly vary when $m$ grows. The lifetimes of 2D rogue events in the short-crested wave fields ($E_{62}$) are characterized by the distributions very similar to the linear case when $m = 2.5$ (Fig. 3c) and $m = 4$ (Fig. 3e). The curves for nonlinear long-crested waves change significantly and somewhat inconsistently in Fig. 3a,c,e; they agree in Fig. 3c, though the tail in the series $A_{12}$ is above the one in the series $E_{12}$ in Fig. 3e. The distributions calculated for the directional wave simulations approximately follow exponential dependencies.

The lifetimes of rogue events in planar wave simulations, $A_0$ and $E_0$, are also plotted in Fig. 3 for the reference. The curves for the series $A_0$ and $E_0$ are close and lie remarkably apart from the directional wave results exhibiting much longer rogue events when $m > 1$ (Fig. 3c,e), while they behave differently if $m = 1$ (Fig. 3a).

Comparing the distributions in the left and right columns in Fig. 3, one may readily see the effect of combining 2D events into 3D events: the latter are noticeably less in numbers (see the total number of events $N_{ev}$ in the legends) and persist significantly longer. The linear cases ($L_{12}$, $L_{62}$) exhibit rather similar dependencies when $m = 2.5$ and $m = 4$ (Fig. 3d,f). Thus, based on the simulated wave ensembles the difference between rogue wave lifetimes in the fields of long-crested and short-crested waves with Gaussian statistics is not found.

The lifetime distributions for long-crested waves with moderate ($A_{12}$) and strong ($E_{12}$) nonlinearity, shown in Fig. 3b for $m = 1$, qualitatively agree with the relation between the dependencies for collinear waves, $A_0$ and $E_0$: lifetimes of rogue events are shorter when waves are steeper and tend to the dependence in the linear limit. The situation is opposite for larger values of $m$ (Fig. 3d,f): nonlinear waves exhibit longer rogue wave lifetimes, especially if the directional spread is small. The distributions of lifetimes of 3D rogue events in the series $A_{12}$, $E_{12}$ differ from the corresponding distributions for 2D events remarkably when $m = 2.5$ and $m = 4$.The difference between the two nonlinear cases of long-crested waves, $A_{12}$ and $E_{12}$, is not prominent according to Fig. 3d,f, though these cases are characterized by much longer rogue events compared to the linear case.

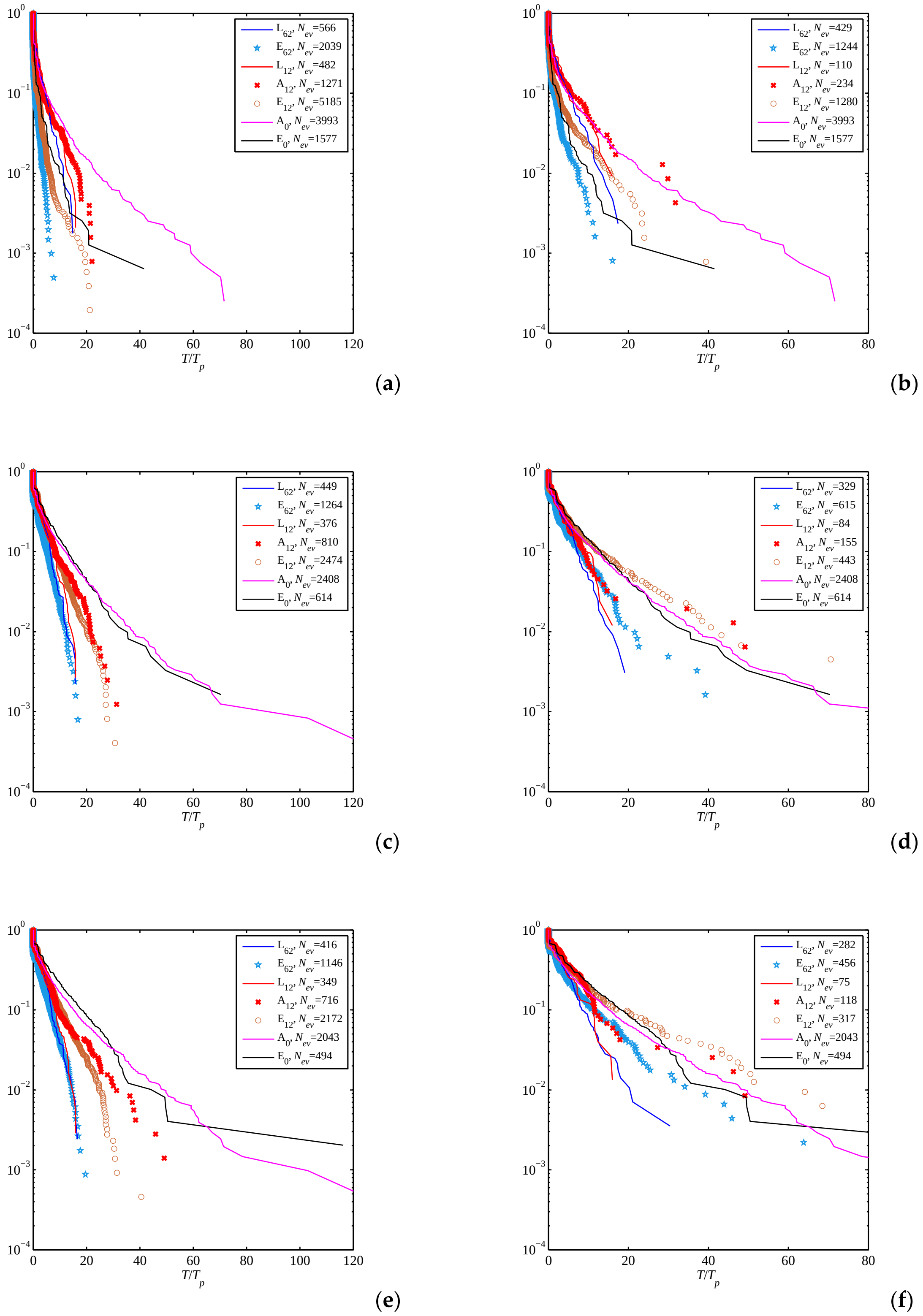

**Figure 3.** Exceedance probability distributions for rogue event lifetimes: (**a**) 2D rogue events, $m = 1$; (**b**) 3D rogue events, $m = 1$; (**c**) 2D rogue events, $m = 2.5$; (**d**) 3D rogue events, $m = 2.5$; (**e**) 2D rogue events, $m = 4$ and (**f**) 3D rogue events, $m = 4$.

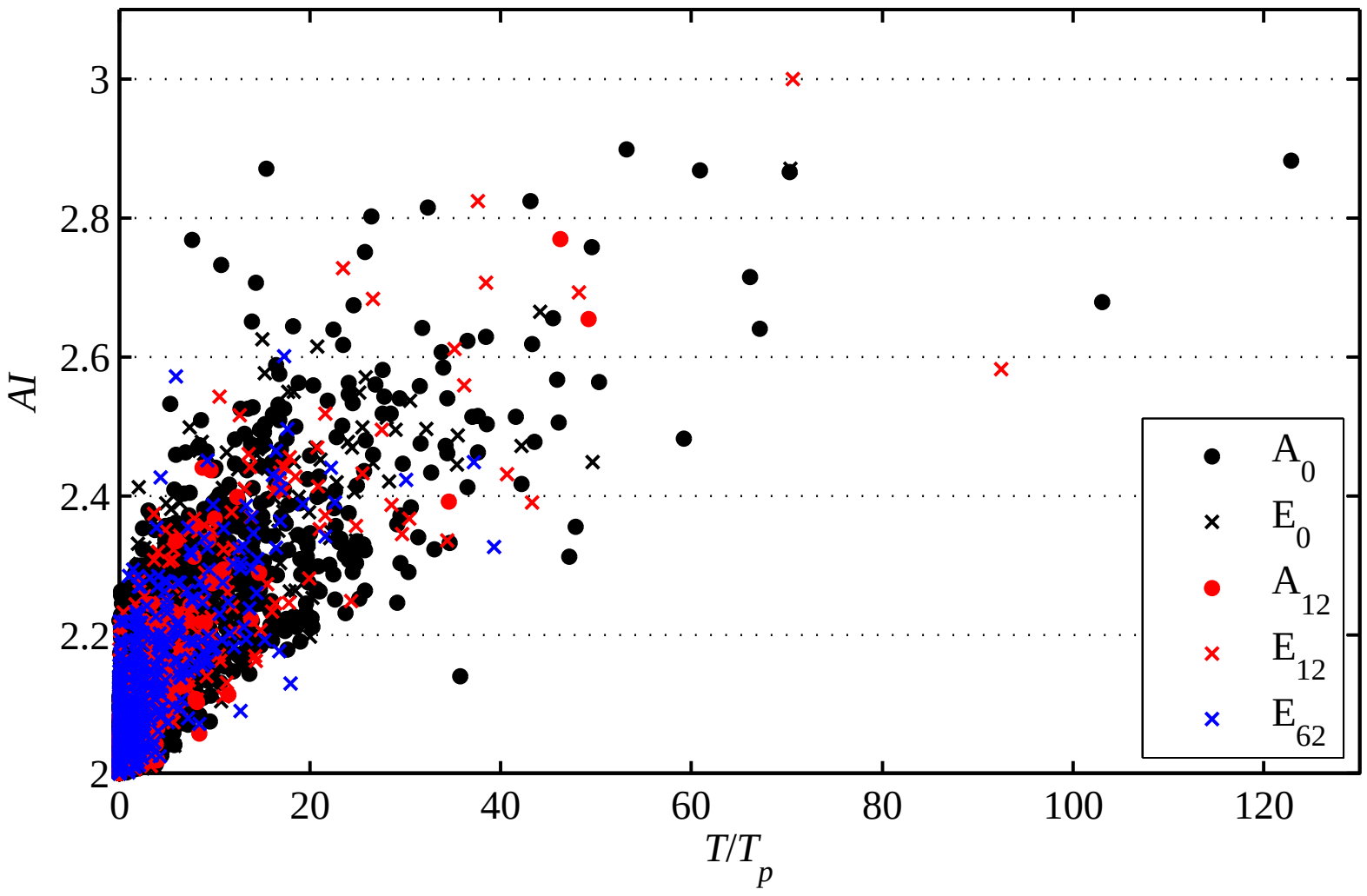

**Figure 4.** Dependence between the rogue event lifetimes $T$ and the amplification factor $AI$, $m$ = 2.5.

According to the results displayed in Fig. 3d,f we may conclude that rogue events persist for longer time in the sea states with narrow angle spectra. Steep waves favor longer events. The lifetime dependencies are close to exponential with slightly heavier tails. The lifetime statistics based on experimental data may be essentially distorted when the wave measurements in lateral positions are absent (i.e., when lifetimes for 2D rogue events are calculated instead of the lifetimes for 3D events).

The relation between the rogue event lifetimes and the amplification factor $AI$ (1) is shown in Fig. 4 for the choice $m$ = 2.5. The scatters for different series exhibit very similar appearances (the linear simulation data are not used), therefore they are plotted in Fig. 4 all together. It follows from the figure that longer-living events are generally characterized by larger amplifications, though a tendency to saturation of $AI$ may be noticed. The maximum wave amplification attained in the performed simulations is about 3. The events in the interval $AI$ =2.8...3 are from the most representative series $A_0$, and also from $E_0$ and $E_{12}$; the corresponding rogue events possess the duration from 15 to 120 wave periods.

A few long-living 3D events found in the simulations $E_{12}$ and $E_{62}$ are shown in Fig. 5,6 and Fig. 7 respectively. Areas of the water surface of the size 20 by 20 dominant wave lengths are shown in the upper parts of the plots. The red color corresponds to elevations, while the blue color – to depressions. The surfaces are centered with respect to the locations of the rogue events; the following system of references moves rightwards with the group velocity of the dominant wave. The crests of the detected rogue waves which belong to the same event are marked with circles; the strokes specify the directions to the deeper adjacent troughs. Longitudinal sections through the locations marked with the circles are shown in the plots below. The values of the maximum amplification index, $AI$, for the displayed rogue waves and of the local significant height, $H_s = 4\sigma$, are given on tops of the panels in Fig. 5-7. Note the variation of the significant wave height estimates $H_s$, which indicates the effect of sampling variability. The displayed instants are chosen within the duration of the rogue event, when the criterion (1) is satisfied.

The events shown in Fig. 5 and Fig. 6 correspond to waves with small directionality (the series $E_{12}$); they last for about 40 wave periods and travel during these time spans for more than 3 km, while the condition (1) is not satisfied continuously during the concerned periods. When $AI$ significantly exceeds the value of two, rogue waves are found in more than one longitudinal cuts (e.g., Fig. 5b,d), hence the lateral size of the rogue wave generally grows with $AI$. The rogue event in Fig. 5 is a part of a long-living slant wave pattern; the characteristic length of the wave front is of the order of 1 km. Generation of rogue waves in slant groups was pointed out in [Ruban, 2011]. A more complicated pattern which contains rogue waves is displayed in Fig. 6 (see in particular Fig. 6a), where essentially directional wave dynamics may be clearly seen. Then the lateral wave structure may be complicated (see Fig. 6b). Rogue waves with different shapes can occur: with large single crests (Fig. 5a,b), sign-changing waves (Fig. 5c, Fig. 6c). Depending on the location, the shapes of rogue waves in Fig. 5d, Fig. 6a,b vary drastically. The deeper trough may be at front or rear slope of the wave; 'holes in the sea' may appear as well.

The long-living event shown in Fig. 7 occurs in a short-crested sea state (the series $E_{62}$) and lasts for shorter time of about 24 dominant wave periods. It belongs to a strongly distorted wave pattern, which is generally shorter in the lateral direction compared to the events shown in Fig. 5,6.

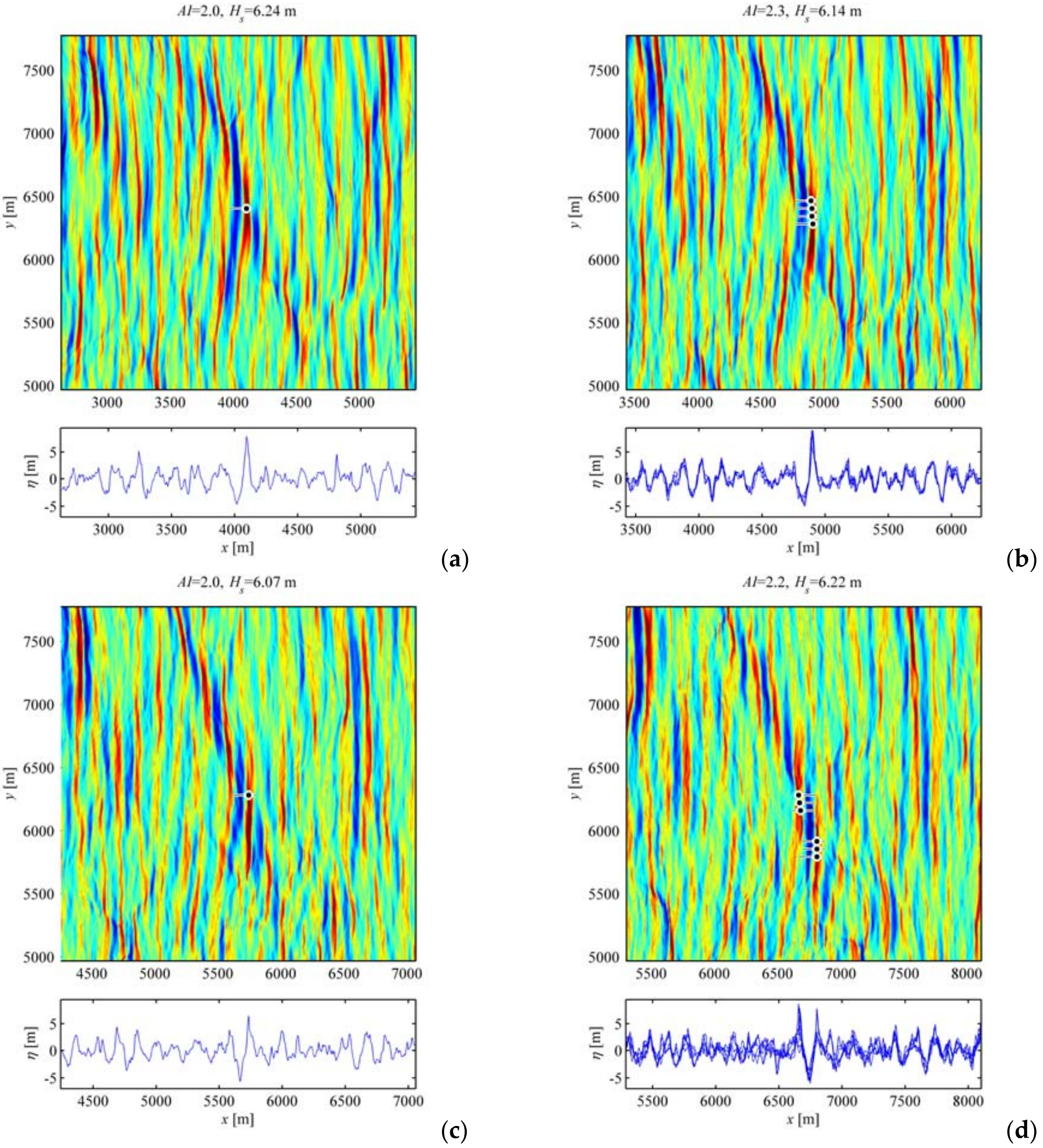

**Figure 5.** Snapshots of the surface in the co-moving frame (upper part) and the corresponding longitudinal wave sections (lower part) during a long-living rogue wave event from the series $E_{12}$. The sections are taken along the points where rogue waves are detected, shown on the surfaces by circles with strokes. The circles show locations of the rogue wave crests; the strokes show directions to the deep troughs. The snapshots correspond to the following instants of time: (**a**) 642.5 s; (**b**) 732.5 s; (**c**) 825 s; and (**d**) 942.5 s.

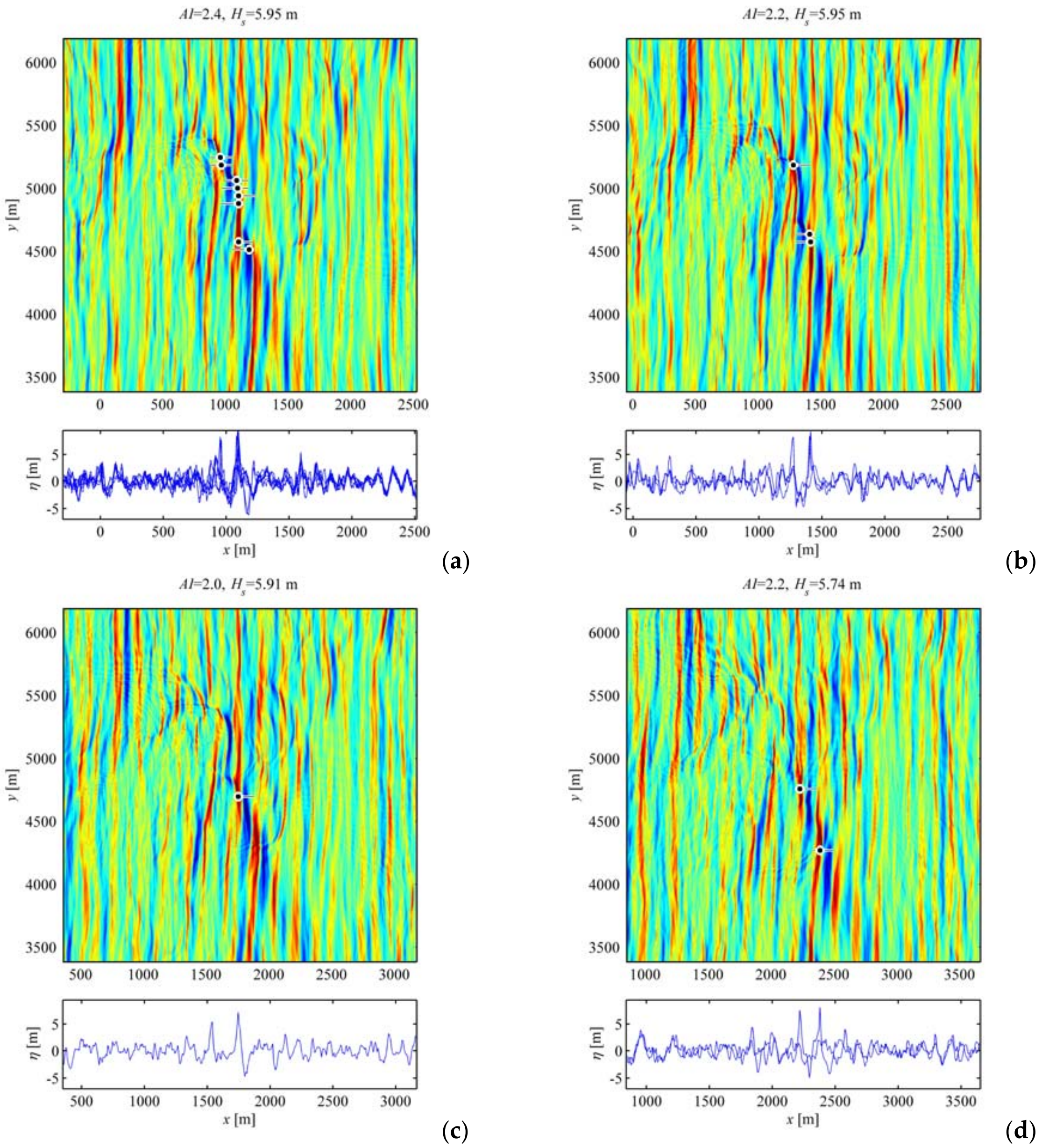

**Figure 6.** Same as in Fig. 5, but a different event is shown from the series $E_{12}$. The snapshots correspond to the following instants of time: (**a**) 529 s; (**b**) 559 s; (**c**) 609 s; and (**d**) 669 s.

## 5. Discussion

In the study of the rogue wave dynamics in irregular seas [Sergeeva & Slunyaev, 2013] we suggested to soften the conventional criterion on rogue waves (1) admitting its temporary violation for a couple of wave periods / wave lengths, and hence reducing the effect of noisy wave perturbations. The fact of surprisingly large lifetimes of rogue wave events, retrieved according to this approach in numerically simulated unidirectional sea states – a few dozens of wave periods – was emphasized in our previous works [Sergeeva & Slunyaev, 2013; Slunyaev et al, 2016; Slunyaev & Kokorina, 2017]. In those publications we conjectured that this result could be a consequence of strictly planar geometry of the simulated wave systems. This guess is partly confirmed in the present work in the sense that when directional waves are concerned, the lifetimes of rogue events found in longitudinal cuts of the wavy surface (we call them 2D rogue events) are much smaller and exhibit rather weak dependence on the particular sea state parameters (such as the wave steepness and broadness of the angle spectrum). The probabilistic distribution of 2D rogue events' lifetimes is close to exponential and does not differ much from the linear theory.

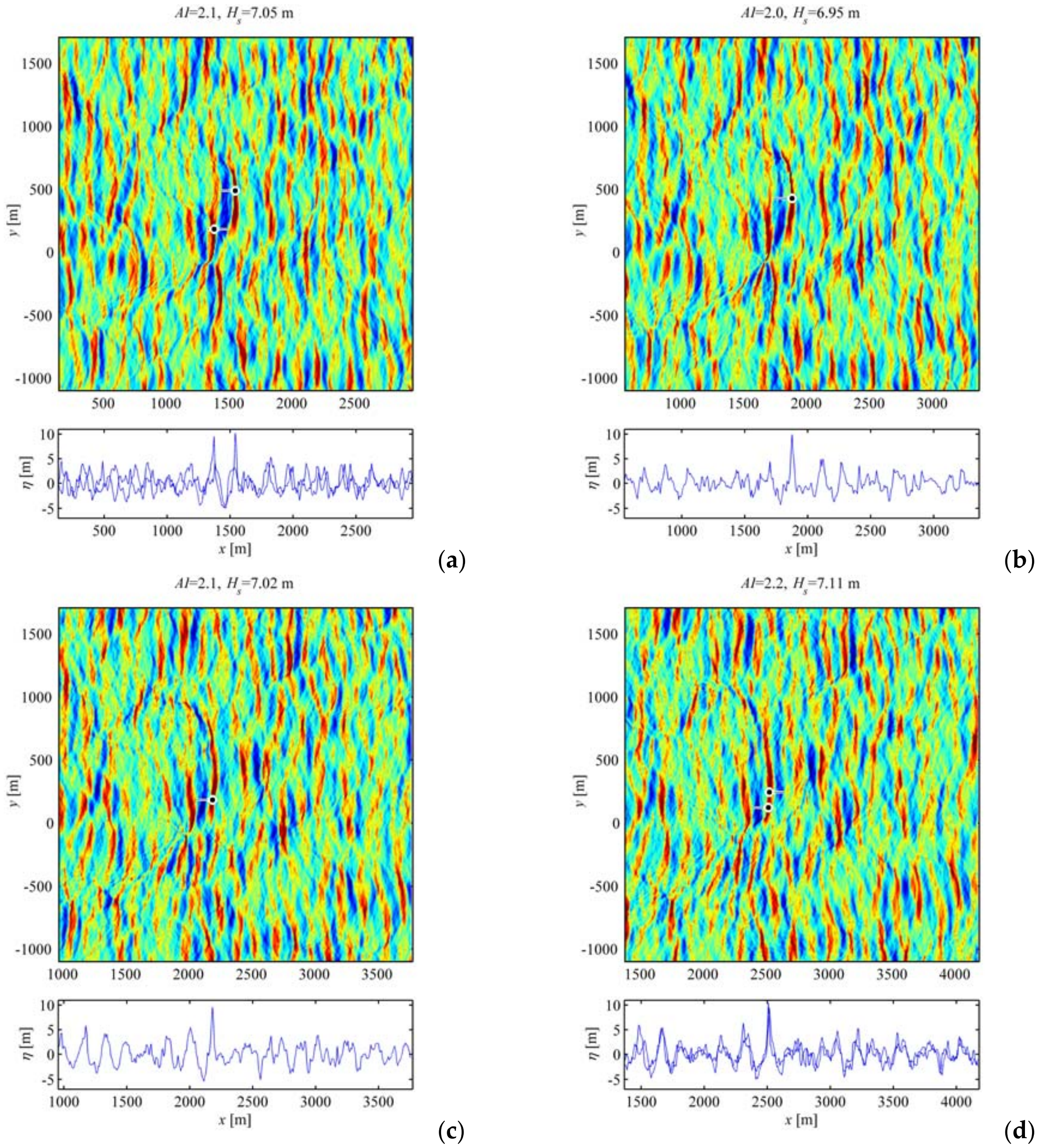

**Figure 7.** Same as in Fig. 6,7, but a long-living rogue wave event from the series $E_{62}$ is shown. The snapshots correspond to the following instants of time: (**a**) 339 s; (**b**) 379 s; (**c**) 419 s; and (**d**) 459 s.

The situation changes when the 2D rogue events are combined into 3D events, taking into account the abnormally high waves in neighboring longitudinal cuts. We consider these structures more physically relevant as they help to trace the generation and evolution of energetic patterns on the two-dimensional sea surface, including possible nonlinear coherent structures. We found that different methods for gathering the abnormally high individual waves to the 3D rogue wave event sets can lead to qualitatively different results in terms of the lifetime distributions, and suggest the approach which enhances stability of the statistical analysis.

In the limit of very small wave amplitudes the distributions for short- and long-crested waves are similar. In rougher sea states the probability of long-living rogue events in long-crested waves grows. This finding agrees with the recent work [Fujimoto et al, 2019]. The probability distribution for steep waves with relatively small directional spread is in a good agreement with the curves for unidirectional waves. Surprisingly, the lifetime distributions for planar and long-crested waves with large and moderate steepness do not differ noticeably; in the performed simulations the maximum rogue wave lifetime in directional seas was about 90 wave periods (for the arrangement parameter $m$ = 2.5). In short-crested rough sea states the lifetimes of rogue waves are shorter.

The remarkable difference between the lifetime distributions for 2D and 3D events in the situations of narrow directional spectra tells that nonlinear JONSWAP wave systems with weak directionality are prone to generation of long-living intense wave patterns with relatively large transverse size. They can live for tens of wave periods and longer, for a few times longer than in the linear theory. Meanwhile the maximum observed lifetime of 3D rogue events in short-crested sea states is at least 20 peak periods, what is more than 3 minutes for 10-sec waves. The probability distributions for lifetimes of 3D events generally follow exponential laws, though the tails in rough conditions of long-crested waves decay slower. No tendency of limiting of the maximum lifetime for even rarer and more extreme events is found, though the saturation of the maximum wave amplification is noticed.

In general, the number of rogue events used for our statistical analysis is not really large, of the order of $10^2 – 10^3$, therefore the formulated statements should be considered as preliminary; more simulations should be performed to make more reliable statements.

The long-living extreme wave patterns may contain large waves of different shapes. In the study we make distinction between the rogue waves with larger front slopes and larger rear slopes. The preliminary analysis confirms that rogue waves with larger rear slopes can dominate under certain conditions. This finding generalizes a similar conclusion of our studies [Sergeeva & Slunyaev, 2013; Slunyaev et al, 2016] to the situation of directional waves. Long-living rogue events in long-crested sea states may be parts of slant patterns as suggested in [Ruban, 2011] or belong to more complicated strongly directional wave structures. In short-crested seas the long-living wave patterns may look intricate.

**Author Contributions:** All parts of the work are performed jointly.

**Funding:** The research was funded by the Russian Foundation for Basic Research (grants Nos. 18-05-80019, 19-55-15005) and by the Fundamental Research Programme "Nonlinear Dynamics" of the Russian Academy of Sciences.

**Conflicts of Interest:** The authors declare no conflict of interest.